# The structure and stability of β-Ta thin films

Aiqin Jiang[1], Trevor A. Tyson[1], Lisa Axe[2], Leszek Gladczuk[3], Marek Sosnowski[3] and Paul Cote[4]

[1]Department of Physics, New Jersey Institute of Technology, Newark, NJ 07102

[2]Department of Civil and Environmental Engineering, New Jersey Institute of Technology, Newark, NJ 07102

[3]Department of Electrical and Computer Engineering, New Jersey Institute of Technology, Newark, NJ 07102

[4]Army Armament Research, Development and Engineering Center, Benét Laboratories, Watervliet, NY 12189

## Abstract

Ta films with tetragonal crystalline structure (β-phase), deposited by magnetron sputtering on different substrates (steel, silicon and silicon dioxide), have been studied. In all cases, very highly preferred (001) orientation was observed in x-ray diffraction (XRD) measurements. All diffraction data revealed two weak reflections corresponding to $d$-spacing of 0.5272 and 0.1777 nm. The presence of the two peaks, attributed to (001) and (003) reflections, indicates that β-Ta films exhibit a high preference for the space group of $P-42_1m$ over $P4_2/mnm$, previously proposed for β-Ta. Differences in relative intensities of (00$l$) reflections, calculated for single crystal β-Ta σ-type Frank-Kasper structure and those measured in the films, are attributed to defects in the films. Molecular dynamics simulations performed on tantalum clusters with six different initial configurations using the embedded-atom-method (EAM) potential revealed the stability of β-Ta, which might explain its growth on many substrates under various deposition conditions.

***Keywords***: XRD; space group; β-tantalum; molecular dynamics; clusters

# 1. Introduction

Tantalum thin films exhibit two crystalline phases, bcc ($\alpha$-phase, the bulk structure of tantalum) and metastable tetragonal $\beta$-phase, which differ in both mechanical and electrical properties. $\alpha$-Ta is considered an attractive coating material in many applications for its high toughness, ductility, low electrical resistivity (15–60$\mu\Omega$.cm) and corrosion resistance [1,2]. Hard and brittle $\beta$-Ta is less desirable in most applications with the exception of thin film resistors because of its high resistivity (170-210 $\mu\Omega$.cm) [3-6]. The structure of Ta films deposited by sputtering is usually the metastable $\beta$-phase or a mixture of the two phases. Special conditions such as increased substrate temperature or local epitaxy are needed to produce $\alpha$-phase Ta [1,2,7]. The mechanism of the preferential growth of $\beta$-phase is not well understood. It is not even clear that this phase has a unique structure as various lattice constants of $\beta$-Ta films [8, 9] and a broad range of resistivity [4-6] have been reported. The study presented in this paper addresses these two issues.

Various crystal structures have been reported for $\beta$-Ta [8 -15] since its discovery. The work by Mosley *et al.* [12] interpreted the x-ray powder diffraction data in terms of a tetragonal unit cell, which is isomorphous with $\beta$-U, with lattice constants *a*=1.0194 / *c*=0.5313 nm and space group *P4$_2$/mnm*. By comparing experimental extended x-ray absorption fine structure (EXAFS) data with theoretically calculated spectra on four candidate structures proposed in the literature, $\beta$-uranium model was also determined to be the most likely structure of $\beta$-Tantalum thin film [13]. A self-hosting $\sigma$-structure with a tetragonal unit cell (lattice constants $a$=1.0211, $c$=0.53064 nm; space group $P\text{-}42_1m$ ) was proposed by Arakcheeva *et al.* [14, 15] on the basis of x-ray diffraction (XRD) study on single crystals of β-Ta produced through electrolytic crystallization. Further investigation of β-Ta single crystals at 293 K and 120 K revealed that the host and guest components of a σ-structure can correspond to different space groups. $P\text{-}42_1m$ has a lower symmetry than $P4_2/mnm$, nevertheless they both belong to the σ-type Frank-Kasper structure [16,17], which is typical for binary intermetallic compounds and β-U.

To determine whether β-Ta thin films show significant preference for $P\text{-}42_1m$ or $P4_2/mnm$ space group, an analysis on the differences and common features of the two structures is presented in terms of x-ray diffraction data. The x-ray diffraction data on four β-Ta thin film samples with thicknesses (~30 μm and ~1 μm), deposited on different substrates (steel, silicon and silicon dioxide) were compared with the calculation based on $P\text{-}42_1m$ coordinates of obtained by Arakcheeva *et al* [14].

To understand the stability of β-Ta phase, we have also performed molecular dynamics simulations on tantalum clusters with sizes up to 5000 atoms. To our knowledge, little work has been done on tantalum clusters [18,19], but some clusters of fcc metals, such as aluminum [20,21], gold [22,23,24] and lead [25,26,27], have been extensively studied. It has been known that atomic clusters of metals are not simply fragments of bulk materials, they can occur in stable arrangements quite different from the bulk crystal structure. With increasing size, the bulk structure begins to dominate due to its lower energy. In our molecular dynamics simulations we have constructed Ta cluster with several different initial configurations and sizes (up to 5000 atoms) and followed their evolution in time at 300 K. The results provide insights into the formation and stability of β-Ta phase in thin films.

## 2. Experimental and Computational Methods

### 2.1 Sample preparation

Tantalum films were deposited by planar magnetron sputtering with argon on

unheated substrates. Four types of β-Ta samples were studied: (A) a 30 µm thick film on

AISI4340 steel substrate; three ~1 µm thick films on AISI4340 steel (B), silicon wafer

(C) and silicon dioxide (D), respectively. All samples were deposited using d.c.

magnetron sputtering source with Ta target of 50 mm diameter and the target – substrate

distance of 50 mm. The thicker sample A was deposited in 1.4 Pa argon gas with a

deposition rate of 3.5 nm/s. Samples B, C and D were deposited in argon gas at 0.7 Pa

with the deposition rate of 1.2 nm/s. More details of sample preparation and the apparatus

can be found in Ref. [28]

### 2.2 X-ray diffraction measurements

For the phase determination of Ta films, XRD measurements with Cu Kα radiation

were conducted using a Philips x-ray diffractometer operated at 45 kV and 40 mA. To

search for (00$l$) β-Ta reflections with odd $l$ for, which are observed in a structure

corresponding to $P\text{-}42_1m$ space group, a long continuous scan of 55 hrs from 15° to 125°

was conducted on all four β-Ta specimens. In addition, control α-Ta samples were measured over very small ranges (about 3°) bracketing small peaks, which were found in the 55 hour long scan on β-Ta. In other diffraction experiments, the effects of film orientation were eliminated by grounding the film of sample C into powder. This was done by removing the film from silicon substrate using adhesive tape, subsequently and cleaned with acetone in an ultrasonic bath. The collected particles were ground in agate mortar and passed through a 200 mesh sieve (particle size < 73.7 μm). The Ta powder containing a small amount of polycrystalline Si was used for XRD measurements. A short scan with a 2θ range from 25° to 50° was performed using a vertical Philips Norelco powder diffractometer (Cu Kα radiation operated at 35 kV/15mA, at CAMET Research, Inc.). The step size and scan rate were 0.04°/step and 25 s/step, respectively. A 12 hour scan in the 2θ region between 15° and 125° with the step size of 0.025°/step and scan rate of 10 s/step was conducted using a horizontal Rigaku 'Geigerflex' powder diffractometer (Cu Kα - radiation operated at 30kV/100mA with RIGAKU RU-200 rotating anode unit, at CAMET Research, Inc.).

### 2.3 Tantalum cluster construction and molecular dynamics simulation

The tantalum clusters were constructed in several forms: (1) rhombic dodecahedrons, fragments of bcc lattice with surfaces cut along [110] planes; (2) two cuboctahedrons: one with equilateral triangular (111) facets, the other with hexagonal (111) facets (both have fcc structure with the surface cut along (111) and (100) planes); (3) noncrystalline icosahedron structure; (4) ball clusters extracted as a sphere from the hcp lattice; (5) ball clusters extracted as a sphere from the σ structure of β-Ta phase with $P\text{-}42_1m$ space group [9, 10]. The first three constructions are shelled structures [29, 30]. Packing of atoms to produce a given polyhedral form leads to sequences of atom numbers with a complete outer shell, which are called magic numbers according to the terminology of Martin [29, 30]. The last two constructions are ball clusters cut from bulk lattice and the clusters in these two groups have incomplete outer shells. Their initial atomic positions are taken to be those of bcc and tetragonal tantalum lattices. Lattice constants with $a$=0.440 nm [4, 31, 32] and $a$=0.311 nm/$c$=0.508 nm were used for fcc and hcp, respectively. The two closed-packed structures were constructed to have the same nearest neighbor distances and density. The icosahedron structure was built to have the nearest neighbouring distance of 0.286 nm which is equivalent to the bulk bcc structure. The size sequences of each group are listed as following:  Rhombic dodecahedrons (bcc) with a sequence of 175, 369, 671, 1105, 1695, 2465, 3439, 4641, 6095; equilateral triangular faceted cuboctahedron (fcc-tri) with a sequence of 147, 309, 561, 923, 1415, 2057, 2869, 3871; hexagonal faceted cuboctahedron (fcc-hex) with a sequence of 201, 586, 1289, 2406, 4033; icosahedron (ico) with the same size sequence as triangular faced cuboctahedron; ball-cut hcp clusters (hcp) with a sequence of 195, 425, 665, 1159, 1555, 2367, 3105, 4331; ball-cut clusters from σ structure of β-Ta (sigma) with a sequence of 129, 505, 1325, 1823, 2761, 3561, 5005.

Molecular dynamics was performed using the molecular-dynamics code IMD originally developed by Stadler [33, 34]. The tantalum clusters are modeled using embedded-atom-method potential constructed by the force-matching method [35]. Simulations were performed in the canonical ensemble at constant temperature of 300 K. The clusters of various structures and sizes were allowed to relax for $1 \times 10^5$ time steps. The time step was chosen as 3.5 fs in all simulations so that the total time of a cluster evolution was 0.35 ns.  The radial distribution function (RDF) was obtained from the end structure as the average over 100 time steps after the cluster was brought to thermal equilibrium.

## 3. Results and Discussion

### 3.1 X-ray diffraction analysis

### 3.1.1 (001) and (003) reflections

The diffraction data collected for 55 hrs on sample A is displayed in Fig. 1. Besides the three peaks referred to as the (002), (004) and (006) reflections with $2\theta = 33.64°$, $70.55°$, $120.04°$ respectively, several very weak peaks are observed at $2\theta = 16.82°$, $38.37°$, $44.62°$, $51.42°$, $107.49°$. The peaks located at $2\theta = 38.37°$ ($d = 0.2346$ nm) and $107.49°$ ($d = 0.0956$ nm) are respectively (110) and (222) reflections from trace amounts of $\alpha$-Ta. Other very small peaks showed in Fig. 1(b) may be due to the substrate and impurities in the coating. The peak at $2\theta = 44.62°$ ($d = \sim0.2031$ nm) was determined to be the (110) reflection from a small area of exposed steel substrate on the sample surface since it exists in both $\alpha$ and $\beta$- Ta coatings on steel but is not present for coatings on Si and $SiO_2$ substrates. This difference may be due to a more uniform coating deposited on Si and $SiO_2$, which are highly polished, smooth surfaces. There are other two weak peaks at $2\theta = 16.82°$ ($d = 0.5272$) and $51.42°$ ($d = 0.1777$ nm) in Fig.1 (a). Short scans (about $3°$ $2\theta$ range) on several other $\beta$-Ta coatings prepared under different conditions reveal that the two small peaks around $2\theta = 16.82°$ and $51.42°$ were observed in the spectra of all tested $\beta$-Ta coatings but not of the control $\alpha$-Ta films.

This peak observed at $2\theta = 16.82°$ has a $d$-spacing value of 0.5272 nm, approximately twice the d-spacing of (002) peak (0.2664 nm). It could be interpreted as the second harmonic of the strong (002) reflection, which is the second order diffraction at half wavelength of the Cu Kα radiation, originating from the continuous Bremsstrahlung background [36]. To check this possibility, an aluminum foil was placed in the diffracted beam path. This did not change the intensity ratio of the two peaks, which would be expected if they were due to diffractions of different wavelengths. It was thus concluded that the peak at $2\theta = 16.82°$ is not from the second harmonic of the (002) reflection. The peak at $2\theta = 51.42°$ could be attributed to the (332) reflection but we assume that this is unlikely for the film with highly preferred orientation indicated by the very strong (002) reflection. An exhaustive search of XRD databases excluded the possibility that the two weak peaks are due to impurities. Lee *et al.* also reported these very weak reflections with β-Ta specimens on steel and glass substrates [37]. This indicates that all β-Ta films exhibit these reflections.

The structure based on the β-U with space group *P4₂/mnm* does not have planes that may cause reflections at $2\theta = 16.82°$. This eliminates the *P4₂/mnm* structure previously proposed for β-Ta [12, 13]. The recently proposed space group *P-42₁m* [14] allows for the reflections at (00*l*) for odd *l*, which are not compatible with *P4₂/mnm*.

The two peaks located at $2\theta = 16.82°$ and $2\theta = 51.42°$ are consistent with reflections from planes with d-spacing approximately twice the spacing of (002) and (006), respectively. Since β-Ta films have strong (002) orientation, the two weak peaks can thus indexed with (001) and (003) reflections. The (005) reflection that may be expected with β-Ta was not observed in our experiments, most likely due to its very low intensity.

**3.1.2 Powder diffraction on sample C**

The diffraction pattern displayed in Fig. 2 was obtained on sample C, which was powdered to eliminate the effects of the preferred film orientation. All very strong peaks are from silicon substrate residue. Reflections (410) and (411) are much broader than the sharp peak (002), indicating reflections from fewer (410) and (411) planes than (002) planes. The difference in the width of the peaks can be explained by the shape of the powdered film particles. The film typically grows in the form of long grains or columns perpendicular to the substrate surface. It breaks up in the milling process into thin filaments elongated in (001) direction, which results in broadening of the XRD peaks, except those that correspond to the reflections from the many planes perpendicular to the column axis. Other expected peaks located between 2θ of 35° and 42°, such as (330) (202), (212) and (331) etc., were not observed because of line broadening and the high background level. Not much more information was gained from the 12 hour scan using the same technique. The lattice parameters obtained from the diffraction measurement were $a$=1.0175 and $c$=0.5312 nm. The 111-Si and 220-Si peaks were used as standards for accurate diffraction angle calibration. The lattice constants are more consistent with that of Mosley $et\ al$. [12] (PDF #25-1280: $a$=1.0194, $c$=0.5313 nm, $P4_2/mnm$) than that of Arakcheeva $et\ al$. [14] ($a$=1.0211, $c$=0.53064 nm, $P\text{-}42_1m$). The position of the strong (002) reflection corresponds to the $c$-axis lattice constant consistent with that reported by Mosley $et\ al$. [12], while the $a$-parameter appears to have decreased by ~0.2 %.

### 3.1.3 Comparison of the diffraction data with calculations

The details of the XRD spectra corresponding to (001), (003) and (002) reflections exhibited by the four samples were plotted in Fig. 3 (a), (b) and (c), respectively. The two weak peaks around 2θ = 16.82° and 51.42°, which are indexed as (001) and (003) reflections, appear in XRD data on all four β-Ta sample. However, the relative intensities of the reflections in the four specimens are different, and the peak positions are also slightly different. The (001) and (003) reflections are much weaker the (002) reflections. The 2θ / d-spacing / relative intensity values measured on the four samples are shown and compared with theoretical calculation in Table 1. The intensities are normalized to (002) reflections in each case.  In the calculations the lattice constants $a$=1.0175 and $c$=0.5312 nm derived from the diffraction measurement on sample C were used. The calculated intensities were corrected for the Lorentz, polarization, multiplicity and temperature factors [14]. The measured intensities of (001), (003) peaks are much lower than the calculated, and (005) peak was not even observed, although it was observed in the β-Ta single-crystal diffraction pattern [14,15]. The experimental peak positions deviated slightly from the calculation. None of the relative experimental intensities is very close to the calculations based on the coordinates obtained by Arakcheeva *et al.* [14].

The structure of thin films often differs from that of the bulk materials. The metastable β-Ta films contain many defects formed in the deposition process, such as inclusion of impurities, strain, dislocations, and sequence faults, which can result in variation of lattice constants and atom coordinates. These effects could account for the differences in relative intensities between the measurement and calculation. Another possible reason for the differences is that β-Ta films and the single crystal may have the same space group as a composite structure but have different space groups for the host or guest components. Arakcheeva *et al.* [15] described recently the self-hosting σ-structure, in which the space group of the host and the guest components and the composite structure of β-Ta single crystal change during a thermal process.

### 3.2 Comparison of two structures: *P-42₁m* and *P4₂/mnm*

The *P-42₁m* and *P4₂/mnm* structures reported for β-Ta are both σ-phase structures belonging to the Frank-Kasper class of tetrahedrally close packed structures [16,17,38]. Fig. 4 shows atomic layers in a typical σ-phase structure. Layer B and C, denoted by hollow balls and gray balls, respectively, represent two primary hexagonal-triangular kagome-tiling nets with three kinds of vertexes ($3636$, $3^2 6^2$, $6^3$) [16,17]. The pseudohexagons in the two primary nets are antisymmetrically superposed. An interlayer (layer A) denoted by black balls and line-shaded balls (behind black balls), with square-triangle net $3^2 434$, centers the superposed hexagons between primary layers. Atoms in the interlayer are all 14-coordinated. The two structures indicated by the two space groups are not strikingly different but they have two chief distinctions. First, the two neighboring A layers in *P-42₁m* structure are not exactly the same. The coordinates of atoms are slightly different. Hence, in Fig. 4 $A_1$ and $A_2$ are used to distinguish the two similar but different layers for *P-42₁m*. Thus *P-42₁m* has an $A_1 BA_2 CA_1 BA_2 C\ldots..$ stacking order while *P4₂/mnm* has an ABACABAC… stacking order. The second difference is in the *z* coordinates of atoms in the B and C layers. In Fig. 4, The A'O'B'C' and AOBC indicate the unit cells of *P-42₁m* and *P4₂/mnm* in the *ab* plane, respectively. In the *c* direction, the centrosymmetry of *P4₂/mnm* results in the perfect planar configuration of atoms in layer B and C at z = 0 and ½, while in *P-42₁m* structure the atoms comprising B and C layers are slightly deviated from the planes at z = ¼ and ¾. In *P-42₁m* the square-triangle nets (layer $A_1$ or $A_2$) are at z~0 and ½ while in *P4₂/mnm* they are at z~ ¼ and ¾.

The occurrence of the (00*l*) reflections with odd *l* in XRD data in *P-42₁m* is due to the two slightly different square-triangle nets (A layers) and the rumpling of the primary nets (B and C layers). The structure factor calculation shows that the *z*-coordinates of atoms determine the relative intensities of (00*l*) lines. These intensities may be affected by differences in atomic *z*-coordinates caused by defects in β-Ta films.

### 3.3 Molecular dynamics simulation analysis

Generally, the potential energy per atom in a cluster decreases with the cluster size because of decreasing percentage of surface atoms. The binding energy, $E_b$, which is defined as  - *V/N* (*V*, potential energy; *N*, number of cluster atoms), is often fitted to the following expansion in N [39, 40]:

$$E_b = a + b\,N^{-1/3} + c\,N^{-2/3} + d\,N^{-1}$$

Similarly, one can fit the potential energy to the above expression, with the coefficients of the opposite sign but the same values as those in the fit to the binding energy.  The average potential energies of relaxed clusters of different sizes with six different initial configurations, together with the fits for the energy of rhombic dodecahedrons (bcc) and ball-cut σ structure (sigma), are plotted against the cluster size N in Fig. 5.  The figure indicates that the relaxed rhombic dodecahedrons with bcc structure are favored energetically over all other relaxed structures. Bcc structure is probably lowest in energy when the cluster size is larger than approximately 200. The sigma sequence has the highest potential compared to other sequences. Almost all points are located between the fit to the sigma sequence and the fit to the bcc sequence. For cluster size smaller than approximately 200, we may expect crossover of potential energies of different structures. For very small clusters, the bulk structure (bcc) may not be the most stable structure as seen for aluminum [21] and lead clusters [27].

The radial distribution functions (RDF) of the initial constructions and the end structures for the largest size clusters in each group are compared in Fig. 6. Inspection of the plots shows that the RDF of relaxed clusters (right column of Fig. 6) for each group, except the sigma phase group, ended with a RDF which is very similar to that of bcc, even though the starting RDF of bcc is quite different than other starting configurations (left column of Fig. 6). Note that the two cuboctahedrons and hcp ball cluster have almost the same initial RDF, because they were all generated from a close-packed bulk structure (12 atoms for the first shell and 6 for the second shell). The RDF plot indicates that the fcc cuboctahedrons, icosahedrons and ball-cut hcp clusters transformed to distorted bcc structures at the end of simulation which have higher potential energies than the relaxed bcc rhombic dodecahedrons (Fig. 5). This suggests that pure tantalum is not stable in fcc or hcp structure at room temperature. The formation of fcc phase of Ta which was reported as being observed in very thin films of tantalum [18, 32, 33] may be caused by incorporation of impurities, or other conditions of film preparation.

The ball-cut σ structure clusters have quite different behavior: instead of transforming to bcc structure, the end structures of all σ structure clusters (including the smallest 129 Ta cluster) maintain the σ structure feature in the simulation (the last two RDF plots in Fig. 6). No phase transformation, which would be indicated by a sharp change of the total energy per atom, occurred in the simulation. The result implies that free-standing β phase clusters with σ-structure are quite stable at room temperature. Moreover, our simulations have shown this structure to be stable up to 400 K (The substrate temperature during deposition of β-Ta is normally below ~400 K [41]). This suggests that the formation of β-phase tantalum in thin films may not be impurity stabilized [42,43,44] or substrate stabilized [45,46], as previously suggested. In addition, a phase transformation to bcc was never observed for the extended β-Ta solid in the simulation work by Klaver [47]. As shown in Fig. 5, the relaxed σ structure clusters have the largest potential energy compared to other groups, which implies that β-phase is a metastable phase of tantalum located at a local minimum on the energy surface. The energy barrier between the local σ-phase minimum and the global minimum, corresponding to the bcc, α-phase, must be sufficiently large for the stability of β-phase Ta.

## 4. Conclusions

XRD studies on β-Ta films deposited by magnetron sputtering on different unheated substrates revealed very strong film orientation corresponding to (001) texture.  Two very weak reflections from planes with spacing around $d = 0.5272$ and $0.1777$ nm, besides three strong reflections (002), (004) and (006), were observed. The analysis of the two small peaks suggests that they may be attributed to (001) and (003) reflections, which would occur in a structure with space group *P-42₁m* but are incompatible with *P4₂/mnm*., previously proposed for β-Ta. Therefore, *P-42₁m* rather than *P4₂/mnm* is suggested as the structure for β-Ta thin films. The comparison of the two σ-type Frank-Kasper structures

*P-42₁m* — render as italic with subscript: $P\text{-}42_1m$ and $P4_2/mnm$ in terms of the nature of layers and major skeletons explains the difference in their diffraction patterns.

A detailed study of the diffraction data reveals differences in the relative intensities of (001), (002) and (003) reflections from four different β-Ta films. There are also significant differences between the calculated relative intensities of (00*l*) reflections and the experimental data.  The differences are attributed to the defects typically present in sputtered films.

Molecular dynamics simulations of tantalum clusters show that among the six constructions studied, the bcc structure has the lowest energy and therefore is most stable for clusters with sizes larger than ~200. The relaxed β phase (σ-structure) tantalum clusters showing higher potential than all other relaxed clusters has also a relatively high energy barrier between the minima corresponding to β and α phases. The stability of β-Ta revealed in the simulation work may explain its growth on many substrates under various deposition conditions.

**ACKNOWLEDGMENTS**

We thank the U.S. Army Sustainable Green Manufacturing Program for the financial support. We would like to thank Dr. L. Keller (CAMET Research, Inc.) for x-ray diffraction measurements and helpful discussions and Dr. N. M. Jisrawi for helpful advice and discussions in molecular dynamics simulation.

# Captions

**Fig. 1** Diffraction patterns with an exposure of 55hrs on sample A (~30 μm β-Ta coating on steel).

**Fig. 2** Diffraction pattern on powder obtained by grinding sample C.

**Fig. 3** Diffraction peaks corresponding to (001), (003) and (002) reflections form four β-Ta samples, as indicated.

**Fig. 4** Construction of a σ-type Frank-Kasper crystal structure obtained by stacking of alternate hexagon-triangle and square-triangle networks of atoms in the (001) plane of the diagram. AOBC and A'O'B'C' indicate the unit cells of *$P4_2/mnm$* and *$P-42_1m$* space groups, respectively.

**Fig. 5** Dependence of the potential energies of relaxed tantalum clusters of different structures on their size. The solid and dashed lines represent fits to bcc and sigma cluster sequence.

**Fig. 6** The RDF of initial configuration (left column) and end structures (right column) of the largest clusters of each group with the cluster sizes indicated in parentheses.

**Table 1.** The 2θ / d-spacings / Relative Intensity comparison between experimental data and theoretical calculation.

| β-Ta samples | 2 θ (°) / d-spacings (nm) / Relative Intensity | | |
|---|---|---|---|
| | 001 | 002 | 003 |
| A | 16.82 / 0.5272 / 0.26 | 33.64 / 0.2664 /100 | 51.42 / 0.1777 / 0.05 |
| B | 16.71 / 0.5305 / 0.07 | 33.66 / 0.2663 / 100 | 51.45 / 0.1776 / 0.02 |
| C | 16.82 / 0.5272 / 1.20 | 33.83 / 0.2650 / 100 | 51.67 / 0.1769 / 0.20 |
| D | 16.71 / 0.5305 / 0.27 | 33.72 / 0.2658 / 100 | 51.64 / 0.1770 / 0.05 |
| Calculation | 16.69 / 0.5312 / 4.09 | 33.75 / 0.2656 / 100 | 51.61 / 0.1771 / 1.69 |

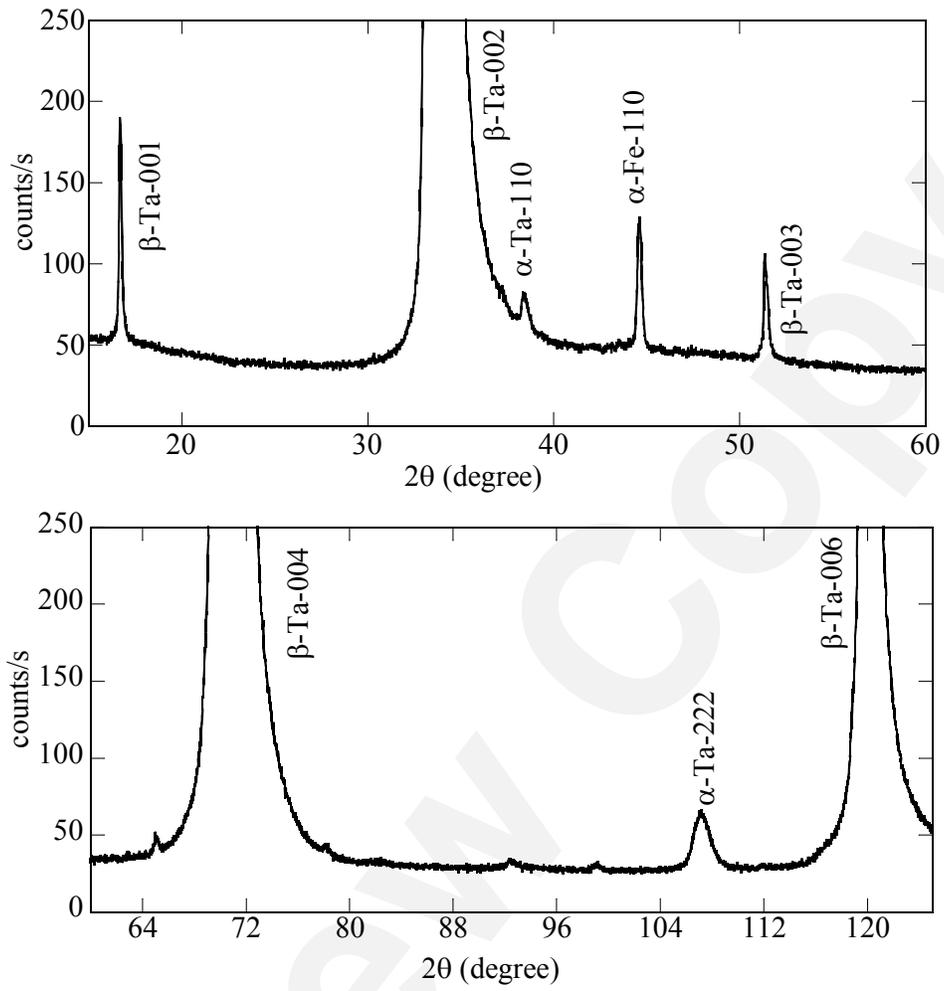

Fig. 1

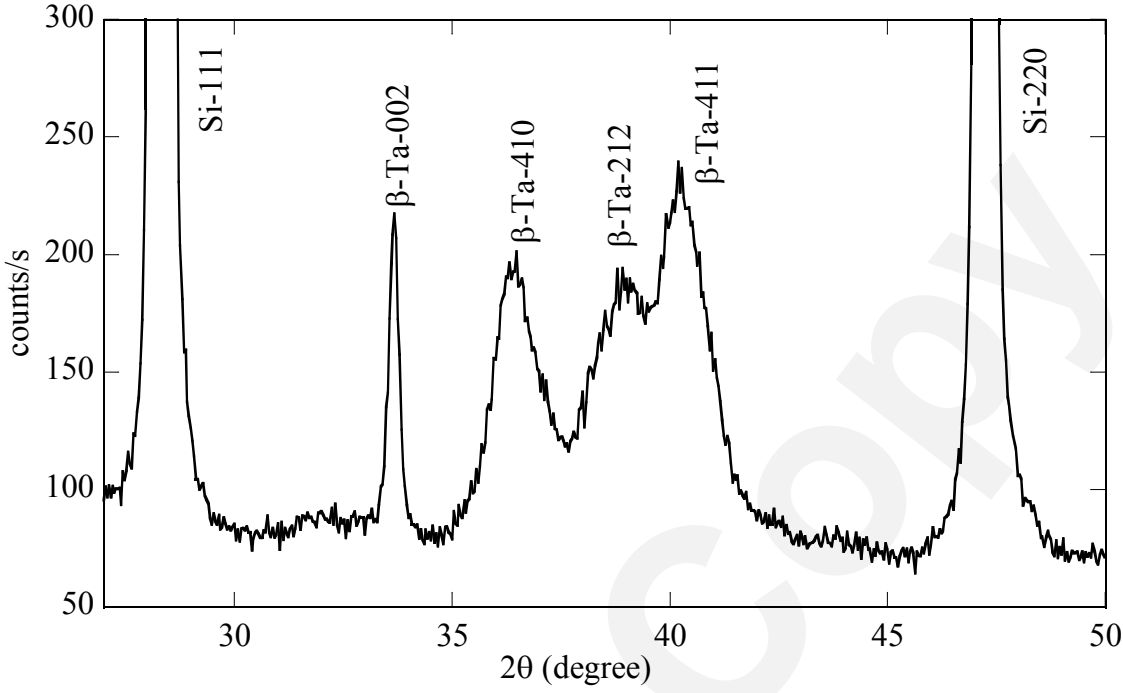

Fig. 2

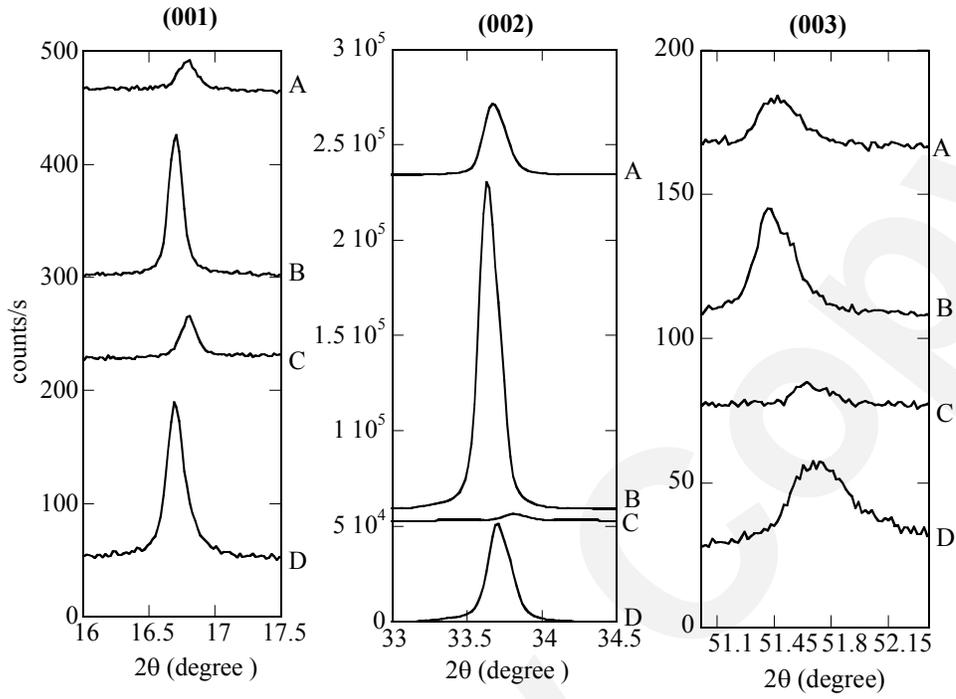

Fig. 3

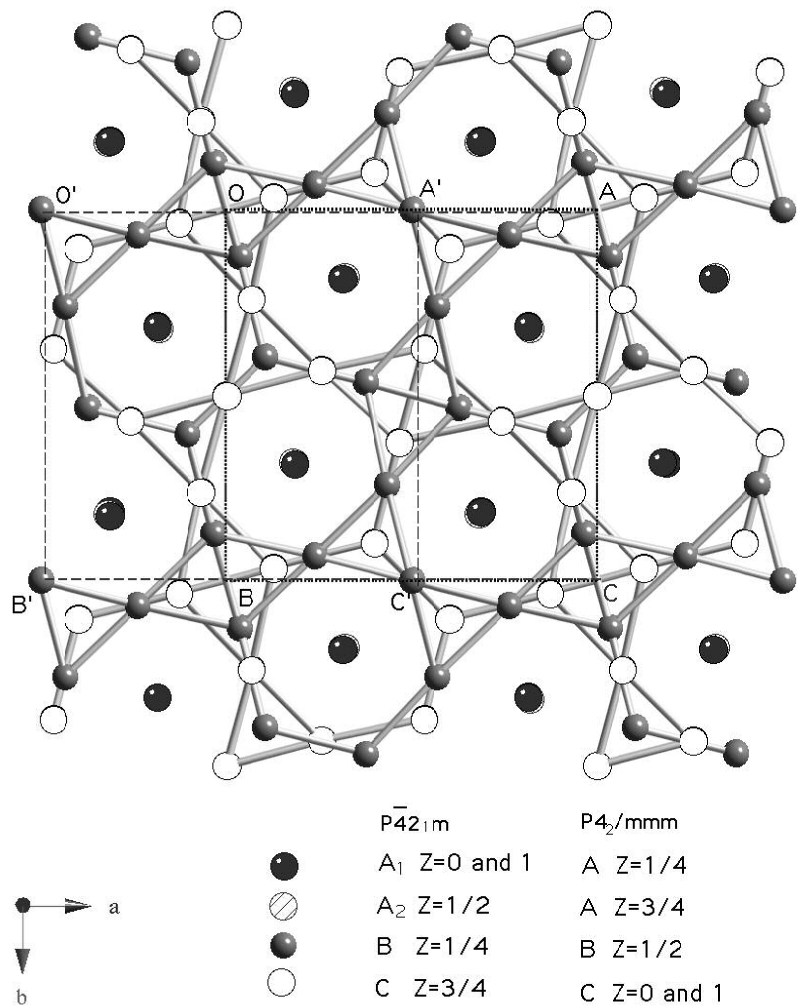

Fig. 4

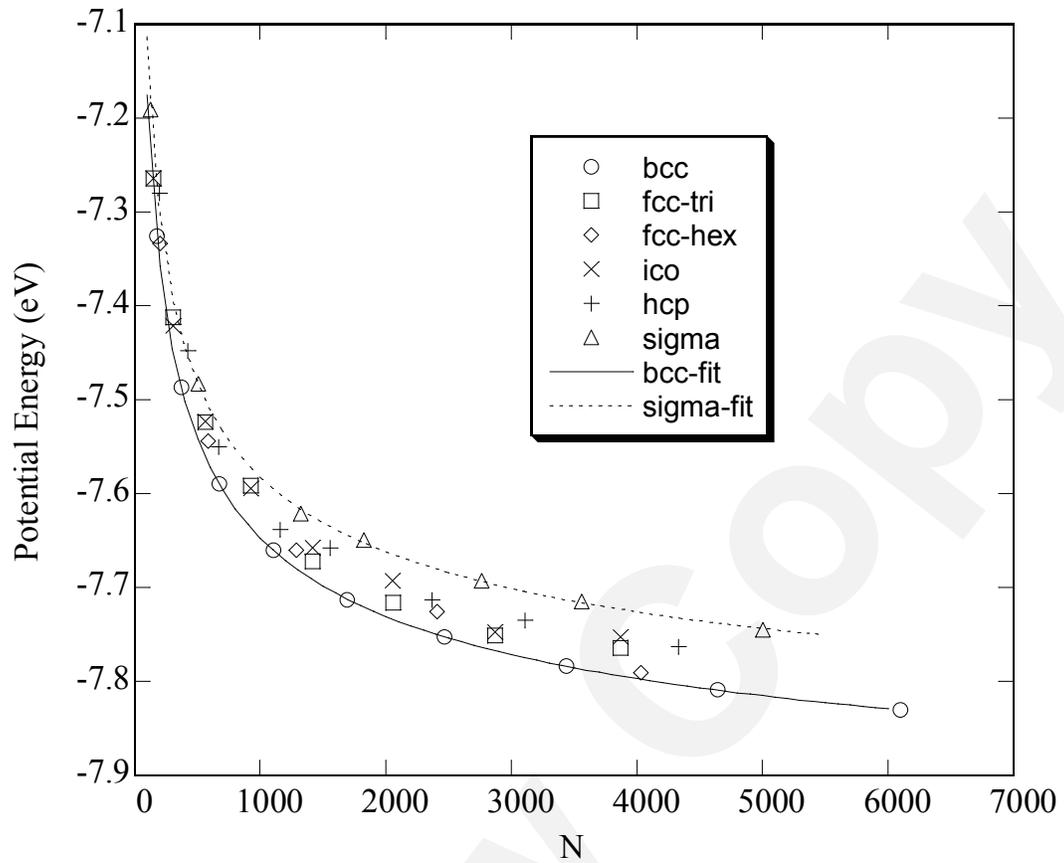

Fig. 5

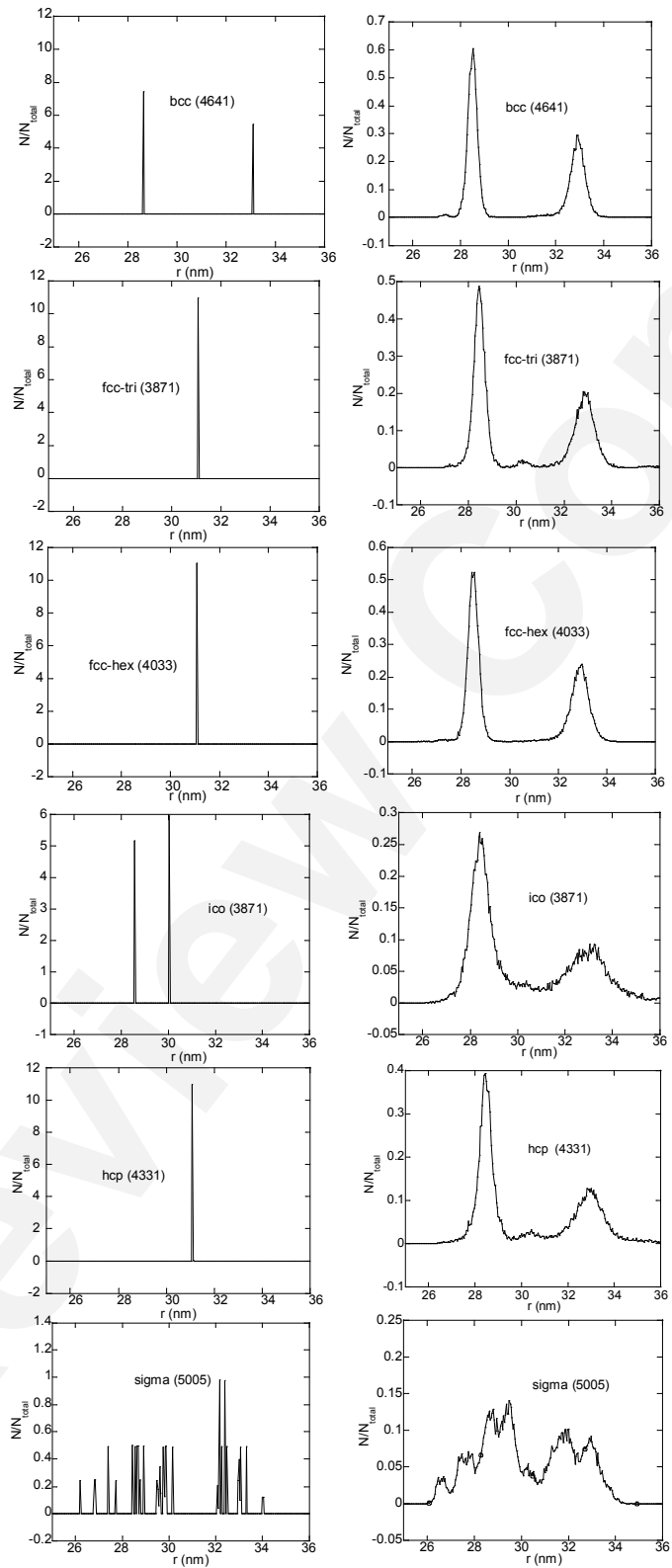

Fig. 6